\documentclass{aastex}
\usepackage{emulateapj5}
\usepackage{amsmath,amssymb,graphicx,natbib}
\newcommand{\Mpc}{{\rm Mpc}}
\newcommand{\hMpc}{\ifmmode{h^{-1}{\rm Mpc}}\else{$h^{-1}{\rm Mpc}$}\fi}
\newcommand{\kms}{\ifmmode{{\rm km}\,{\rm s}^{-1}}\else{km\,s$^-1$}\fi}
\newcommand{\dt}{{{\rm d}t\,}}
\newcommand{\dchi}{\!\!{\rm d}\chi\,}
\newcommand{\m}{{\rm m}}
\newcommand{\rb}{{\rm b}}
\newcommand{\re}{{\rm e}}
\newcommand{\bk}{{\bf k}}
\newcommand{\bxi}{{\boldsymbol{\xi}}}
\newcommand{\btheta}{{\boldsymbol{\theta}}}
\newcommand{\kaco}[1]{\left\langle{#1}\right\rangle}

\begin{document}

\title{Effects of weak lensing on the topology of CMB maps}

\author{
Jens Schmalzing\altaffilmark{1,2},
Masahiro Takada, and
Toshifumi Futamase
}
\affil{
Astronomical Institute,
Graduate School of Science,
Tohoku University,
Sendai 980-8578,
Japan.
}
\affil{
jens,takada,tof@astr.tohoku.ac.jp
}
\altaffiltext{1}{
Teoretisk Astrofysik Center,
Juliane Maries Vej 30,
DK-2100 K\o benhavn \O,
Denmark.
}
\altaffiltext{2}{
Ludwig-Maximilians-Universit\"at,
Theresienstra\ss e 37,
D-80333 M\"unchen,
Germany.
}

\begin{abstract}
We investigate the non-Gaussian signatures in Cosmic Microwave
Background (CMB) maps induced by the intervening large-scale structure
through weak lensing.  In order to measure the deviation from the
Gaussian behavior of the intrinsic temperature anisotropies, we use a
family of three morphological descriptors, the so-called Minkowski
functionals.  We show analytically how these quantities depend on the
temperature threshold, and compare the results to numerical
experiments including the instrumental effects of Planck.  Minkowski
functionals can directly measure the statistical properties of the
displacement field and hence provide useful constraints on large-scale
structure formation in the past.
\end{abstract}

\keywords{
Methods:~statistical --- 
cosmic~microwave~background ---
dark~matter --- 
gravitational~lensing ---
large-scale~structure~of~Universe
}

\section{Introduction}

The temperature anisotropy in the CMB is a powerful probe of the
content and nature of our Universe.  Most inflationary scenarios
{\citep{sato1981,guth1981}} predict that the temperature fluctuation
field obeys Gaussian statistics {\citep{guth1985}}, so all its
statistical properties can be accurately predicted based on the
Gaussian theory {\citep{bardeen1986:bbks,bond1987}}.  However,
gravitational lensing by the matter inhomogeneities between the last
scattering surface and us imprints non-Gaussian signatures on the CMB.
These signatures directly probe the mass distribution up to very high
redshift, and their measurement could greatly help to constrain
cosmological parameters.

Various statistical methods have been used to investigate lensing of
CMB maps.  The effect on the power spectrum $C_\ell$ itself is rather
small {\citep{seljak1996:gravitational}}.  The probability density
function (PDF) of peak ellipticities of the lensed temperature field
is in principle sensitive to the lensing signatures on scales below
$10'$, but the finite beam size of detectors tends to circularize the
deformed ellipticities again {\citep{bernardeau1998}}.  Using the
correlation between ellipticities of the lensed CMB map and distant
galaxies proves more robust against the beam smearing effect
{\citep{waerbeke2000:lensing}}.  {\citet{takada2000:gravitational}}
measure lensing signatures on scales around $75'$ with the two-point
correlation function of hotspots.

In this {\em Letter}, we suggest to detect weak lensing signatures in
CMB maps with the Minkowski functionals {\citep{minkowski1903}}.  To
illustrate that the method is useful at all, we perform quantitative
analyses on numerical experiments {\citep{takada2000:detectability}}.
Section~\ref{sec:numerical} describes the production of the CMB maps
used in our work.  In Section~\ref{sec:minkowski}, we summarize some
of the properties of the Minkowski functionals and apply them to the
maps.  Section~\ref{sec:outlook} provides an outlook.  The Appendix
briefly derives the dependence of the Minkowski functionals of a
lensed CMB map on the temperature threshold.

\section{Numerical experiments}
\label{sec:numerical}

We consider two currently fashionable cosmologies, namely Cold Dark
Matter variants with $\Omega_\m$=1 and $h$=0.5 (SCDM) and
$\Omega_\m$=0.3, $\Omega_\lambda$=0.7, and $h$=0.7 ($\Lambda$CDM).
$\Omega_\m$ and $\Omega_\lambda$ are the energy densities of matter
and vacuum relative to the critical density, respectively, and $h$ is
the Hubble constant in units of $100\kms\Mpc^{-1}$.  The baryon
density is chosen as $\Omega_\rb h^2=0.019$ from recent Big Bang
Nucleosynthesis results {\citep{burles1998}}.  We use a
Harrison-Zel'dovich initial power spectrum and the transfer function
of {\citet{bardeen1986:bbks}} with the shape parameter form
{\citep{sugiyama1995}}.  After that, we can only vary the
normalization, or equivalently $\sigma_8$, the mass fluctuation in a
sphere of radius 8\hMpc.

For each model, we calculate the $C_\ell$ with CMBFAST
{\citep{seljak1996:lineofsight}} and compute 30 realizations of the
CMB on an $80\degr\times80\degr$ square divided into $2'$ pixels.

The lensing effect can be treated as a mapping of the primary CMB
anisotropies.  When we observe the CMB temperature $\psi$ in a certain
direction $\btheta$ on the sky, we actually see the intrinsic field
$\varphi$ at a different position $\btheta+\bxi$ on the last
scattering surface:
\begin{equation}
\psi(\btheta)=\varphi(\btheta+\bxi(\btheta)).
\label{eq:lensing}
\end{equation}
To obtain the lensing displacement field $\bxi$, we first generate the
convergence field $\kappa$.  Its power spectrum $P_\kappa(\ell)$ is
the matter power spectrum $P_{\delta}(k)$ projected along the line of
sight:
\begin{multline}
P_\kappa(\ell)
=
\tfrac{9}{4}H_0^4\Omega_\m\\
\times\int_0^{\chi_{\text{rec}}}\dchi
a^{-2}\left(\frac{r(\chi_{\text{rec}}-\chi)}{r(\chi_{\text{rec}})}\right)^2
P_{\delta}\left(k=\frac{\ell}{r(\chi)},\chi\right).
\end{multline}
$\chi$ and $r(\chi)$ denote the comoving radial distance and the
corresponding comoving angular diameter distance, and
$\chi_{\text{rec}}$ is the radial distance to the last scattering
surface.  From $\kappa$ we compute the displacement field $\bxi$ in
Fourier space through $\bxi(\bk)=2i\bk{k^{-2}}\kappa(\bk)$ and
transform to real space.  According to Equation~(\ref{eq:lensing}),
the displacement field only distorts the locations of the values in
the primordial map.  So we interpolate the original temperature values
with a cloud-in-cell procedure to obtain the lensed temperature map on
a regular grid.  Finally, we mimic an observation through Planck
{\citep{mandolesi1995}} by adding a Gaussian beam smearing and white
noise.  The beam FWHM and the noise level per FHWM pixel are chosen as
$5.5'$ and $4.3\times10^{-6}$, respectively {\citep{efstathiou1999}}.

\begin{figure}
\includegraphics[width=.48\linewidth]{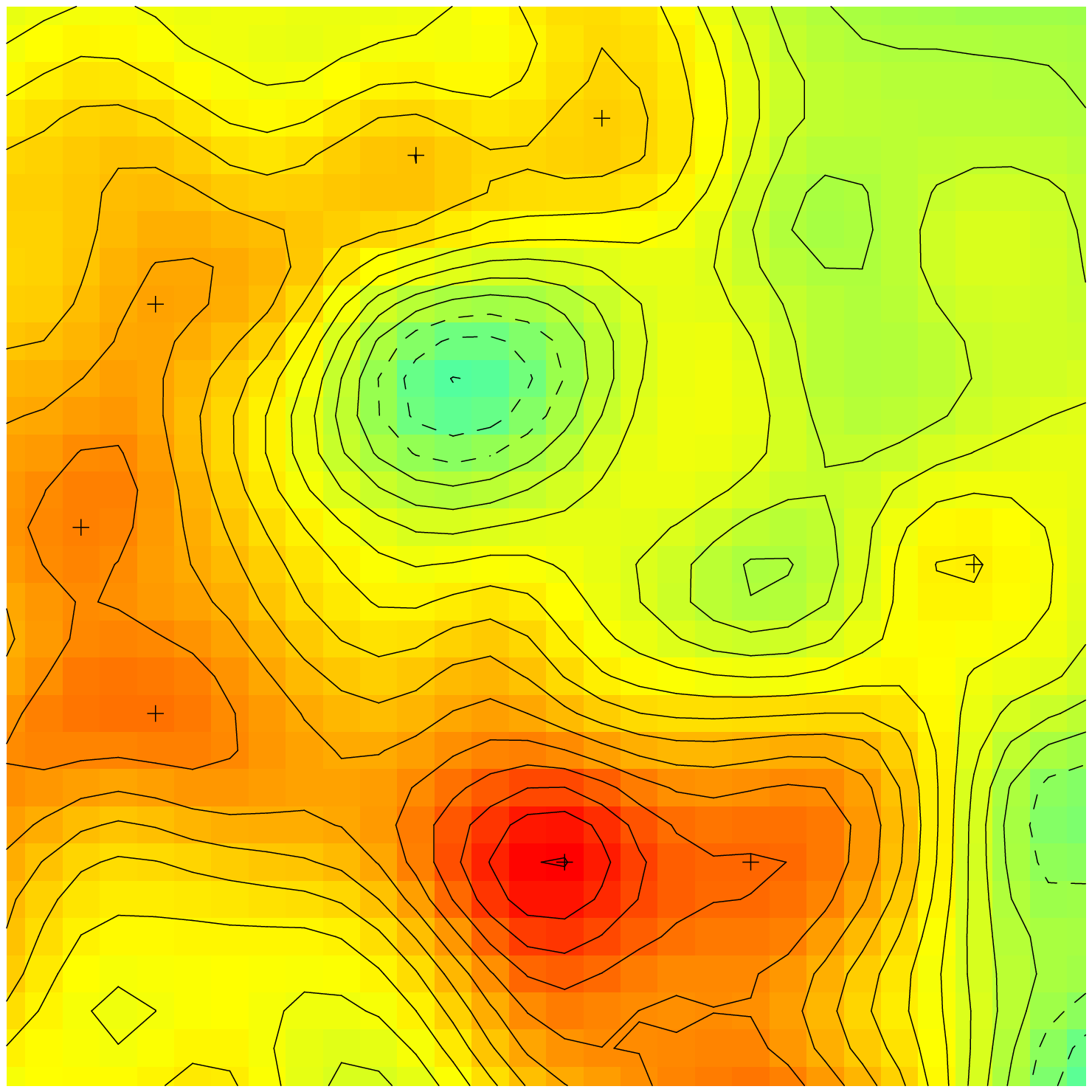} \hfill
\includegraphics[width=.48\linewidth]{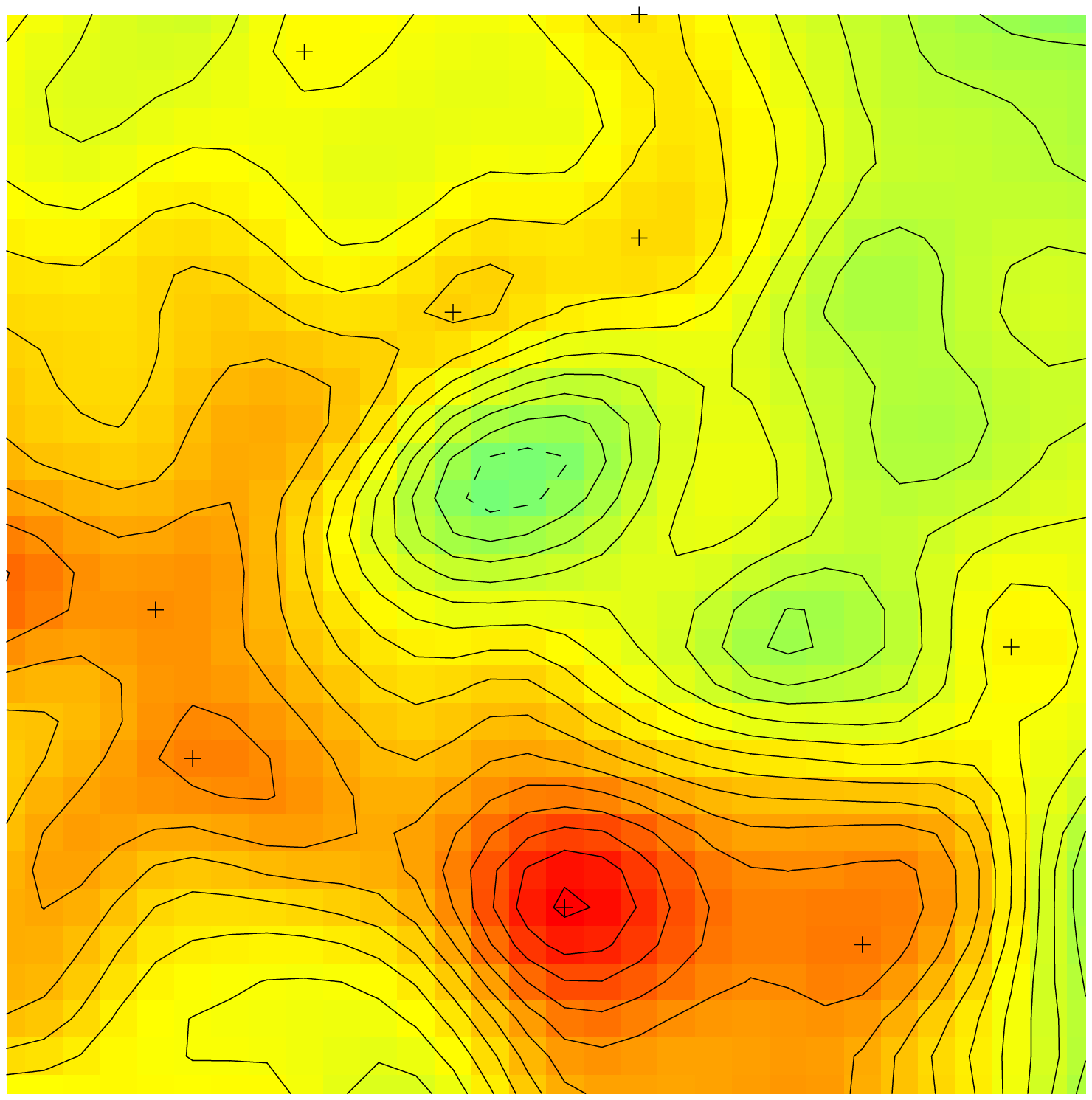}
\caption{
\label{fig:example}
The same patch of the microwave sky seen without (left) and with
(right) the lensing effect by the large-scale structure of a Standard
Cold Dark Matter model normalized with $\sigma_8=1.5$.  The
temperature of the CMB is reflected in the temperature of the colors,
and contours are drawn at intervals of 0.5 times the variance.  Both
the deformation of individual peaks and the distortion of the relative
positions of the peaks are clearly visible.  }
\end{figure}

As an example, Figure~\ref{fig:example} shows a map with and without
lensing.

\section{Integral geometry for CMB maps}
\label{sec:minkowski}

Minkowski functionals were formally introduced into cosmology by
{\citet{mecke1994}}.  Although the whole family of these geometric
descriptors was already mentioned in {\citep{coles1988}},
{\citet{schmalzing1998}} first applied them to CMB maps.
{\citet{bernardeau1997}} suggested Minkowski functionals for detecting
weak lensing signatures in the CMB.  For a continuous field, we use
the Minkowski functionals to describe the geometry of its excursion
sets, that is the area above a given temperature threshold.  In two
dimensions, there are three Minkowski functionals which correspond to
well-known geometrical quantities, namely the area $v_0$, the
circumference $v_1$, and the Euler characteristic $v_2$.

The unlensed CMB maps obey Gaussian statistics.  Therefore, their
statistical properties can be completely characterized by the power
spectrum $C_\ell$.  Following {\citet{tomita1990}}, we parametrize the
Minkowski functionals of the isotemperature contour at the normalized
threshold $\nu$ with $\tau$, the variance of the field's first
derivative\footnote{$\Phi(x)=(2/\sqrt{\pi})\int_0^x\dt\exp(-t^2)$ is
the Gaussian error function.  We assume that the field is normalized
to unit variance.}
\begin{equation}
\begin{split}
v_0^{\text{Gauss}}(\nu)
&=
\frac{1}{2}-\frac{1}{2}\Phi\left(\frac{\nu}{\sqrt{2}}\right),
\qquad\\
v_1^{\text{Gauss}}(\nu)
&=
\frac{\sqrt{\tau}}{8}\re^{-\nu^2/2},
\qquad\\
v_2^{\text{Gauss}}(\nu)
&=
\frac{\tau}{\sqrt{8\pi^3}}\nu\re^{-\nu^2/2}.
\label{eq:analytical}
\end{split}
\end{equation}

The Appendix shows that the {\em shapes} of the Minkowski functional
curves of the lensed field remain the same as in the Gaussian case.
So the non-Gaussianity manifests itself only in the {\em
normalization} of the curves.  Let us denote the constants of
proportionality from Equations~(\ref{eq:v1}) and (\ref{eq:v2}) with
$\alpha_1$ and $\alpha_2$, respectively.
Equation~(\ref{eq:analytical}) yields
\begin{equation}
\alpha_1^{\text{Gauss}}=\frac{\sqrt{\tau}}{8},
\qquad
\alpha_2^{\text{Gauss}}=\frac{\tau}{\sqrt{8\pi^3}}.
\end{equation}
Therefore, for the Gaussian random field the quantity
\begin{equation}
\alpha=\left(\tfrac{8}{\pi}\right)^{3/2}\frac{\alpha_1^2}{\alpha_2}
\label{eq:gaussianity}
\end{equation}
equals unity.  As we shall see, lensing leads to deviations from
unity.

\begin{figure}
\includegraphics[width=.48\linewidth]{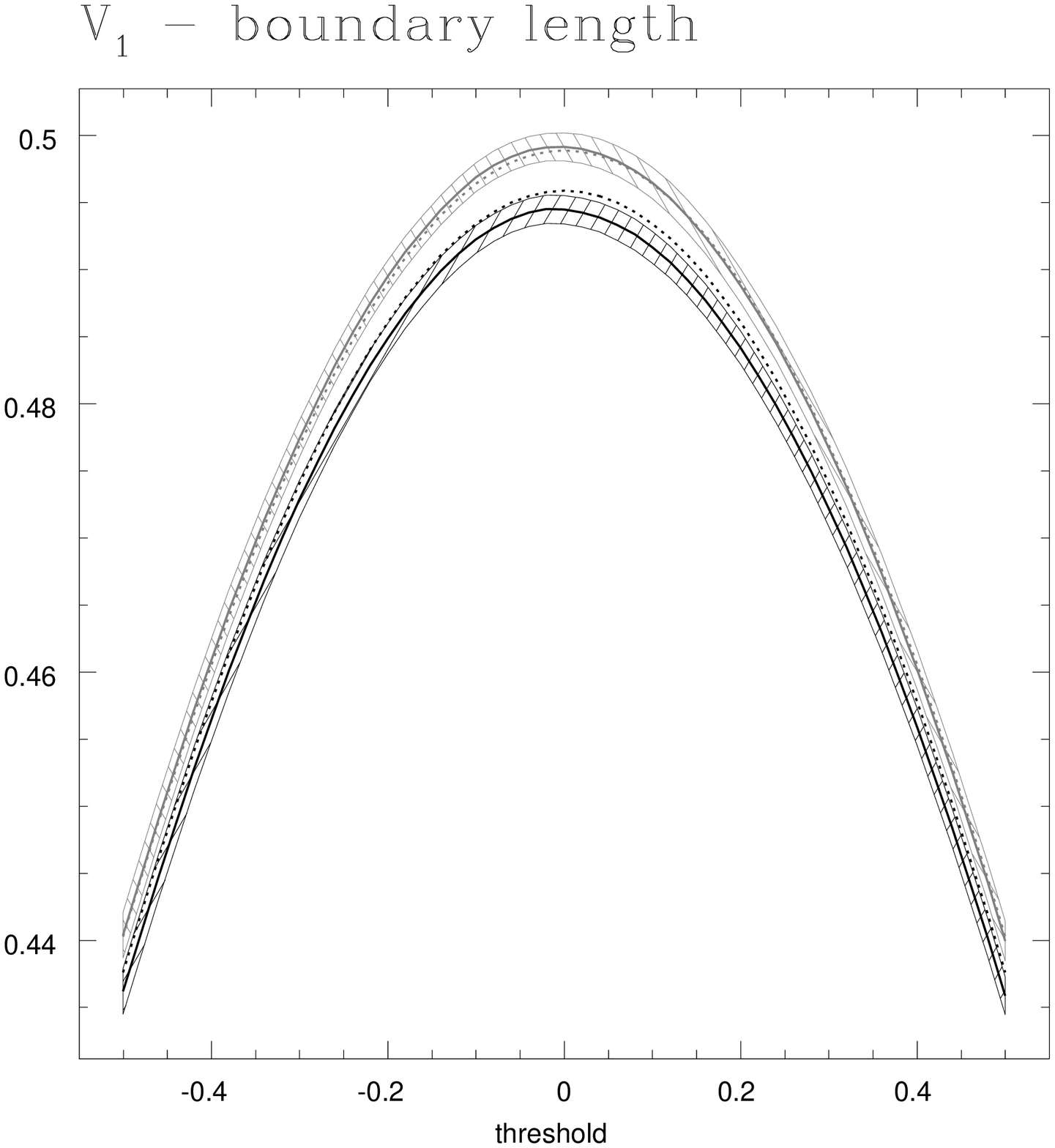} \hfill
\includegraphics[width=.48\linewidth]{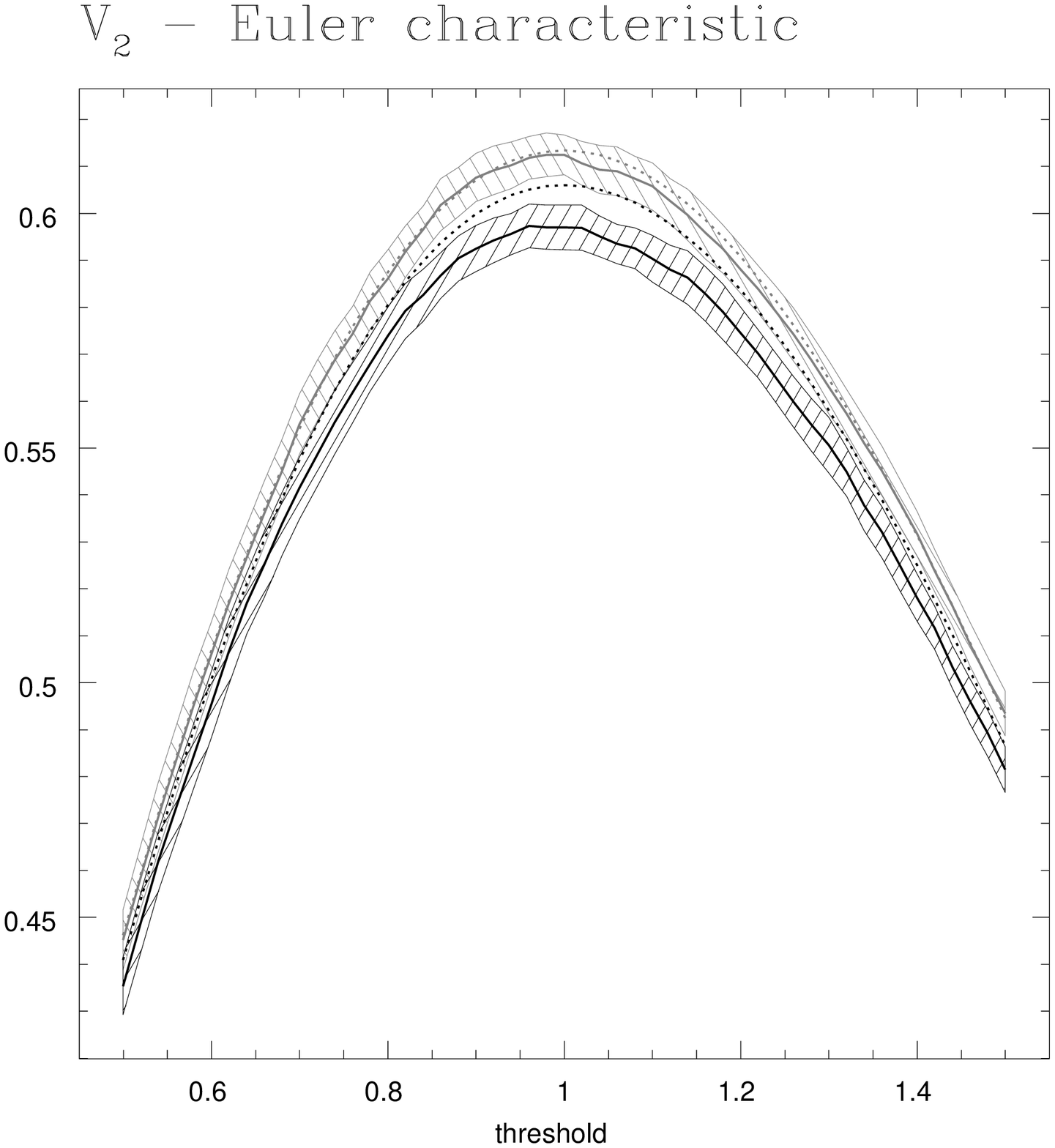}
\caption{
\label{fig:minkowski}
Minkowski functionals for the lensed and unlensed maps observed by the
Planck satellite.  Results for the unlensed maps are shown in grey,
while black lines indicate the Minkowski functionals of the lensed
maps.  For both, the average and standard deviation are shown in solid
lines, while the expectation values for a Gaussian random field with
the same two-point characteristics are displayed in dashed lines.
Note that we only show the region around the maxima of the Minkowski
functional curves, where the differences are visibly significant.  }
\end{figure}

Even with high resolution and large sky coverage, these changes are
easily overcome by the cosmic error.  Therefore, accurate evaluation
of the Minkowski functionals is crucial.  We use three independent
methods and take care that they produce compatible results.  One of
them uses contour integration to evaluate the circumference and the
Morse theorem {\citep{morse1969}} to determine the Euler
characteristic (see e.g.\ {\citealt{novikov1999:minkowski}}).  Two
more elaborate methods {\citep{schmalzing1997:beyond}} evaluate the
Minkowski functionals via Crofton's formula {\citep{crofton1968}}, and
by averaging over invariants formed from derivatives, respectively.
Figure~\ref{fig:minkowski} shows the Minkowski functionals of one of
the models.

For each realization, we determine the non-Gaussianity parameter
$\alpha$ by fitting the measured Minkowski functionals to their
expected Gaussian shapes.  Since each of our realizations covers only
15\% of the sky, while for the Planck satellite a sky coverage of 60\%
and more is expected, we reduce our estimated variances by a factor of
two.  It is also worth mentioning that we assume Gaussianity of the
convergence field on scales of 5' and above.  This is not exactly true
{\citep{jain2000}}, so we expect an even stronger signal from a
refined analysis with a convergence field calculated from $N$--body
simulations.

Table~\ref{tab:gaussianity} summarizes our results.  As expected,
neither of the unlensed models deviates significantly from
Gaussianity.  For the maps that include the weak lensing effect, the
average $\alpha$ is different from one.  In all but one case, this
difference is significant, and in two out of the four investigated
models with lensing, the significance level is above 95\%.

\section{Discussion and Outlook}
\label{sec:outlook}

We have measured the weak lensing effect of large-scale structure on
the observed temperature anisotropies of the CMB with Minkowski
functionals.  Numerical simulations have shown that the effect can be
significant when observed with the experimental specifications of
Planck.  It remains to be seen whether the Minkowski functionals can
directly measure any characteristics of large-scale structure.  Since
they are sensitive to smoothing, we expect that varying the smoothing
scale can reveal information on the convergence field on different
scales.  Most importantly, however, we have proven that this method
can measure non-Gaussian signatures induced by weak lensing at all.

\section*{Acknowledgements}

JS and MT acknowledge support by the JSPS.  This work was supported by
Danmarks Grundforskningsfond through its support for TAC.

\bibliography{bibliography}

\begin{thebibliography}{27}
\expandafter\ifx\csname natexlab\endcsname\relax\def\natexlab#1{#1}\fi

\bibitem[{Adler(1981)}]{adler1981}
Adler, R.~J. 1981, The geometry of random fields (Chichester: John Wiley \&
  Sons)

\bibitem[{Bardeen {et~al.}(1986)Bardeen, Bond, Kaiser, \&
  Szalay}]{bardeen1986:bbks}
Bardeen, J.~M., Bond, J.~R., Kaiser, N., \& Szalay, A.~S. 1986,
  \providecommand{\apj}{Ap.\ J.}{\apj}, 304, 15

\bibitem[{Bernardeau(1997)}]{bernardeau1997}
Bernardeau, F. 1997, \providecommand{\aanda}{Astron.\ Astrophys.}{\aanda}, 324,
  15

\bibitem[{Bernardeau(1998)}]{bernardeau1998}
---. 1998, \providecommand{\aanda}{Astron.\ Astrophys.}{\aanda}, 338, 767

\bibitem[{Bond \& Efstathiou(1987)}]{bond1987}
Bond, J.~R. \& Efstathiou, G. 1987, \providecommand{\mnras}{Mon.\ Not.\ Roy.\
  Astron.\ Soc.}{\mnras}, 226, 655

\bibitem[{Burles \& Tytler(1998)}]{burles1998}
Burles, S. \& Tytler, D. 1998, \providecommand{\apj}{Ap.\ J.}{\apj}, 499, 699

\bibitem[{Coles(1988)}]{coles1988}
Coles, P. 1988, \providecommand{\mnras}{Mon.\ Not.\ Roy.\ Astron.\
  Soc.}{\mnras}, 234, 509

\bibitem[{Crofton(1868)}]{crofton1968}
Crofton, M.~W. 1868, Phil.\ Trans.\ Roy.\ Soc.\ London, 158, 181

\bibitem[{Efstathiou \& Bond(1999)}]{efstathiou1999}
Efstathiou, G. \& Bond, J.~R. 1999, \providecommand{\mnras}{Mon.\ Not.\ Roy.\
  Astron.\ Soc.}{\mnras}, 304, 75

\bibitem[{Guth(1981)}]{guth1981}
Guth, A.~H. 1981, Phys.\ Rev.\ D, 23, 347

\bibitem[{Guth \& Pi(1985)}]{guth1985}
Guth, A.~H. \& Pi, S.-Y. 1985, Phys.\ Rev.\ Lett., 49, 1110

\bibitem[{Jain {et~al.}(2000)Jain, Seljak, \& White}]{jain2000}
Jain, B., Seljak, U., \& White, S. 2000, \providecommand{\apj}{Ap.\ J.}{\apj},
  530, 547

\bibitem[{Mandolesi {et~al.}(1995)Mandolesi, Bersanelli, Cesarsky, Danese,
  Efstathiou, Griffin, Lamarre, {Norgaard-Nielsen}, Pace, Puget, Raisanen,
  Smoot, Tauber, \& Volonte}]{mandolesi1995}
Mandolesi, N., Bersanelli, M., Cesarsky, C., Danese, L., Efstathiou, G.,
  Griffin, M., Lamarre, J.~M., {Norgaard-Nielsen}, H.~U., Pace, O., Puget,
  J.~L., Raisanen, A., Smoot, G.~F., Tauber, J., \& Volonte, S. 1995, Planetary
  and Space Science, 43, 1459

\bibitem[{Mecke {et~al.}(1994)Mecke, Buchert, \& Wagner}]{mecke1994}
Mecke, K.~R., Buchert, T., \& Wagner, H. 1994, \providecommand{\aanda}{Astron.\
  Astrophys.}{\aanda}, 288, 697

\bibitem[{Minkowski(1903)}]{minkowski1903}
Minkowski, H. 1903, Mathematische Annalen, 57, 447, in German

\bibitem[{Morse \& Cairns(1969)}]{morse1969}
Morse, M. \& Cairns, S.~S. 1969, Critical point theory in global analysis and
  differential topology (New York and London: Academic Press)

\bibitem[{Novikov {et~al.}(1999)Novikov, Feldman, \&
  Shandarin}]{novikov1999:minkowski}
Novikov, D.~I., Feldman, H.~A., \& Shandarin, S.~F. 1999, Int. J. Mod. Phys.,
  D8, 291

\bibitem[{Sato(1981)}]{sato1981}
Sato, K. 1981, \providecommand{\mnras}{Mon.\ Not.\ Roy.\ Astron.\
  Soc.}{\mnras}, 195, 467

\bibitem[{Schmalzing \& Buchert(1997)}]{schmalzing1997:beyond}
Schmalzing, J. \& Buchert, T. 1997, \providecommand{\apj}{Ap.\ J.}{\apj}, 482,
  L1

\bibitem[{Schmalzing \& G{\'o}rski(1998)}]{schmalzing1998}
Schmalzing, J. \& G{\'o}rski, K.~M. 1998, \providecommand{\mnras}{Mon.\ Not.\
  Roy.\ Astron.\ Soc.}{\mnras}, 297, 355

\bibitem[{Seljak(1996)}]{seljak1996:gravitational}
Seljak, U. 1996, \providecommand{\apj}{Ap.\ J.}{\apj}, 436, 1

\bibitem[{Seljak \& Zaldarriaga(1996)}]{seljak1996:lineofsight}
Seljak, U. \& Zaldarriaga, M. 1996, \providecommand{\apj}{Ap.\ J.}{\apj}, 469,
  437

\bibitem[{Sugiyama(1995)}]{sugiyama1995}
Sugiyama, N. 1995, \providecommand{\apjs}{Ap.\ J.\ Suppl.}{\apjs}, 100, 281

\bibitem[{Takada \& Futamase(2000)}]{takada2000:detectability}
Takada, M. \& Futamase, T. 2000, accepted for publication in ApJ,
  astro-ph/0008377

\bibitem[{Takada {et~al.}(2000)Takada, Komatsu, \&
  Futamase}]{takada2000:gravitational}
Takada, M., Komatsu, E., \& Futamase, T. 2000, \providecommand{\apj}{Ap.\
  J.}{\apj}, 533, L83

\bibitem[{Tomita(1990)}]{tomita1990}
Tomita, H. 1990, in Formation, dynamics and statistics of patterns, ed.
  K.~Kawasaki, M.~Suzuki, \& A.~Onuki, Vol.~1 (World Scientific), 113--157

\bibitem[{{van Waerbeke} {et~al.}(2000){van Waerbeke}, Bernardeau, \&
  Benabed}]{waerbeke2000:lensing}
{van Waerbeke}, L., Bernardeau, F., \& Benabed, K. 2000,
  \providecommand{\apj}{Ap.\ J.}{\apj}, 540, 14

\end{thebibliography}

\appendix

\section{Analytical treatment}

Consider a scalar random field $\psi$ in two dimensions, e.g.\ the
observed CMB temperature anisotropy.  It can be related to a scalar
Gaussian random field $\varphi$, the intrinsic temperature anisotropy,
through the vector-valued random field $\bxi$, the displacement, by
Equation~(\ref{eq:lensing}).

It is well-known that the PDF of the temperature field, and hence the
zeroth Minkowski functional $v_0$, which is just the integrated PDF,
does not change at all under lensing.  We write the other two
Minkowski functionals as spatial averages over invariants formed from
the field's derivatives\footnote{Indices following a comma denote a
spatial derivative.  } {\citep{schmalzing1998}}:
\begin{equation}
v_1(\nu) =
\frac{\pi}{4}\kaco{\delta(\psi-\nu)\sqrt{\psi_{,1}^2+\psi_{,2}^2}}
,\qquad
v_2(\nu) = 
\frac{1}{2\pi}\kaco{\delta(\psi-\nu)
\frac
{2\psi_{,1}\psi_{,2}\psi_{,12}-\psi_{,1}^2\psi_{,22}-\psi_{,2}^2\psi_{,11}}
{\psi_{,1}^2+\psi_{,2}^2}}.
\label{eq:invariant}
\end{equation}

In order to evaluate these two averages, we need to express the first-
and second-order derivatives of $\psi$ in terms of the fields
$\varphi$ and $\bxi$.  Straightforward differentiation
yields\footnote{Summation over pairwise indices is implied
throughout.}
\begin{equation}
\psi_{,i} = \varphi_{,i}+\varphi_{,k}\xi_{k,i},
\qquad
\psi_{,ij} = \varphi_{,ij}+\varphi_{,ik}\xi_{k,j}+\varphi_{,jk}\xi_{k,i}+\varphi_{,kl}\xi_{k,i}\xi_{l,j}+\varphi_{,k}\xi_{k,ij}.
\label{eq:derivative}
\end{equation}
For the circumference $v_1$, Equation~(\ref{eq:invariant}) involves
only the first derivatives of the observed field $\psi$ and therefore,
by Equation~(\ref{eq:derivative}), only the first derivatives of the
intrinsic field $\varphi$.  Since these are independent of the value
$\varphi$ itself {\citep{adler1981}}, and $\bxi$ is of course
independent of the field $\varphi$, the average splits neatly into two
factors:
\begin{equation}
v_1(\nu)
=
\frac{\pi}{4}\kaco{\delta(\psi-\nu)}
\kaco{\sqrt{\psi_{,1}^2+\psi_{,2}^2}}
=
\frac{\pi}{4}\frac{\re^{-\nu^2/2}}{\sqrt{2\pi}}
\kaco{\sqrt{\psi_{,1}^2+\psi_{,2}^2}}
.
\label{eq:length}
\end{equation}
The remaining average is independent of $\nu$, so the curve has the
Gaussian shape:
\begin{equation}
v_1(\nu)\propto\re^{-\nu^2/2}.
\label{eq:v1}
\end{equation}
Turning to the Euler characteristic $v_2$, we observe that
Equation~(\ref{eq:invariant}) expressed in terms of the fields
$\varphi$ and $\bxi$ depends linearly on the second derivatives
$\varphi_{,ij}$ and $\xi_{k,ij}$.  Therefore, $v_2$ depends on the
threshold $\nu$ only through
\begin{equation}
\kaco{\delta(\varphi-\nu)\varphi_{,ij}}=
-\tau\nu\frac{\re^{-\nu^2/2}}{\sqrt{2\pi}}\delta_{ij}
\qquad\text{and}\qquad
\kaco{\delta(\varphi-\nu)\xi_{k,ij}}=0.
\label{eq:average}
\end{equation}
The remainder of the average in Equation~(\ref{eq:invariant}) again
produces factors that do not depend on the threshold $\nu$.  So the
curve $v_2(\nu)$ also has the shape expected for the Gaussian case:
\begin{equation}
v_2(\nu)\propto\nu\re^{-\nu^2/2}.
\label{eq:v2}
\end{equation}

\setcounter{section}{0}

\begin{deluxetable}{ccccc}
\tablecaption{ The ``non-Gaussianity parameter'' $\alpha$ from
Equation~\protect~(\ref{eq:gaussianity}).  For a Gaussian random
field, $\alpha$ should be unity, while deviations from one are
expected for a weakly lensed CMB sky.  $\alpha$ and $\Delta\alpha$ are
the average and standard error over all realizations of each model.
\label{tab:gaussianity}
}
\tablewidth{0pt}
\tablehead{
\colhead{model} & 
\colhead{experiment} & 
\colhead{$\sigma_8$} & 
\colhead{$\alpha$} & 
\colhead{$\Delta\alpha$}
}
\startdata
 $\Lambda$CDM & Planck & no lensing & 1.00019  & 0.00182 \\
 $\Lambda$CDM & Planck & 1.0        & 1.00099  & 0.00153 \\
 $\Lambda$CDM & Planck & 1.5        & 1.00209  & 0.00184 \\
 $\Lambda$CDM & Planck & 2.0        & 1.00502  & 0.00186 \\
 SCDM         & Planck & no lensing & 0.99912  & 0.00171 \\
 SCDM         & Planck & 1.5        & 1.00355  & 0.00183 \\
\enddata
\end{deluxetable}

\end{document}